\documentclass{ifacconf}

\usepackage{graphicx}      
\usepackage{natbib}        
\usepackage{amsmath,graphicx}
\usepackage{ amssymb}
\usepackage{mathrsfs}
\usepackage[hyphens]{url}

\usepackage{mathtools}
\mathtoolsset{showonlyrefs}

\newtheorem{Def}{Definition}

\newtheorem{Rema}{Remark}

\newcommand{\erfc}{\operatorname{erfc}}

\DeclareMathOperator{\v1}{V1}

\def\cF{{\mathcal F}}

\def\N{{\mathbb N}}    
    
\def\R{{\mathbb R}}

    \def\cS{{\mathcal S}}                     \def\cF{{\mathcal F}}

\begin{document}

	\begin{frontmatter}
		
		\title{Cortical origins of MacKay-type visual illusions: A case for the non-linearity\thanksref{footnoteinfo}} 
		
		\thanks[footnoteinfo]{This work has been supported by the ANR-20-CE48-0003.
			The first author was supported by a grant from the ``Fondation CFM pour la Recherche''.}
		
		\author[First]{Cyprien Tamekue}.
		\author[First]{Dario Prandi}.
		\author[First]{Yacine Chitour}.
		
		\address[First]{Universit\'{e} Paris-Saclay, CNRS, CentraleSup\'{e}lec, Laboratoire des Signaux et Systèmes, 91190, Gif-sur-Yvette, France (e-mail: \{cyprien.tamekue, dario.prandi, yacine.chitour\}@centralesupelec.fr).}
		
		\begin{abstract}                
			To study the interaction between retinal stimulation by redundant geometrical patterns and the cortical response in the primary visual cortex ($\v1$), we focus on the MacKay effect (Nature, 1957) and Billock and Tsou's experiments (PNAS, 2007). We use a controllability approach to describe these phenomena starting from a classical biological model of neuronal field equations with a non-linear response function. The external input containing a localised control function is interpreted as a cortical representation of the static visual stimuli used in these experiments. We prove that while the MacKay effect is essentially a linear phenomenon (i.e., the nonlinear nature of the activation does not play any role in its reproduction), the phenomena reported by Billock and Tsou are wholly nonlinear and depend strongly on the shape of the nonlinearity used to model the response function.
		\end{abstract}
		
		
	\end{frontmatter}
	
	\section{Introduction}
	
\begin{figure}
		\centering
		\includegraphics[width=.35\linewidth]{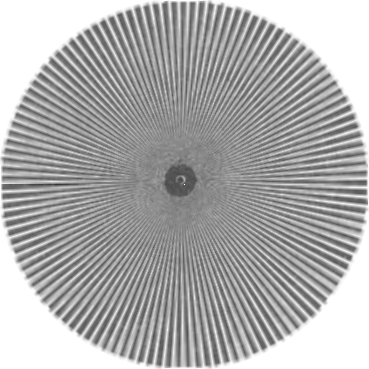}\hspace{1em}
		\includegraphics[width=.35\linewidth]{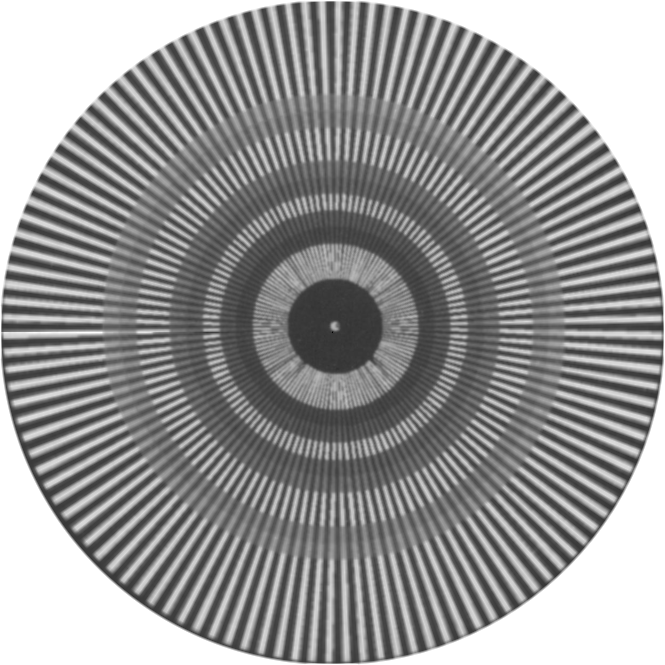}
		\caption{MacKay effect: the presentation of the stimulus to the \emph{left} (``MacKay rays'') induces an illusory perception of the image on the \emph{right}. Adapted from \cite{mackay1957} and \cite{zeki1993}.}
		\label{fig:mackay}
	\end{figure}

	%
	%
	%
	In many situations, humans perceive an illusory component that is not physically present in a visual stimulus.
	\cite{helmholtz1867} is probably the first to be interested in the visual effect induced by the presentation of a pattern consisting of black-and-white zones. In particular, he related the perception of rotating darker and brighter radial zones after viewing a pattern consisting of black and white concentric rings to the fluctuation of eye accommodation.
	In this direction, \cite{mackay1957} reported striking after-effect of visual stimulation by regular geometrical patterns with highly redundant information (see Fig.~\ref{fig:mackay} for the so-called ``MacKay rays'') and attributed the phenomena to some part of the visual cortex, which might profit from such redundancy. In these experiments, an illusory contour  consisting of a pattern of white and black concentric rings (tunnel pattern) is evoked by all observers as the after-image induced by a pattern consisting of white and black fan shape (funnel pattern) with high redundant information in the fovea (the centre of the visual field).
	%
	Due to the retino-cortical map\footnote{Let $(r,\theta)\in[0,\infty)\times[0,2\pi)$ denote polar coordinates in the visual field (or in the retina) and $(x_1,x_2)\in\R^2$ Cartesian coordinates in $\v1$. The retino-cortical map (see, e.g., \cite{tamekue2022} and references within) is given by 
		\begin{equation*}
			r e^{i\theta}  \mapsto (x_1,x_2):=\left( \log r, \theta \right).
	\end{equation*}} between the retina and the primary visual cortex ($\v1$, henceforth), the after-images in $\v1$ are superimposed patterns consisting of orthogonal horizontal and vertical stripes. 
	This indicates that neuronal response in $\v1$ tends to favour directions at a right angle to the visual stimulus. 
	
\begin{figure}
		\centering
		\includegraphics[width=.35\linewidth]{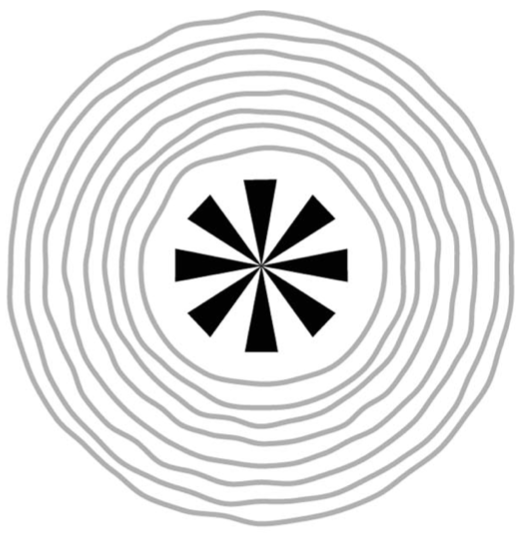}\hspace{1em}
		\includegraphics[width=.35\linewidth]{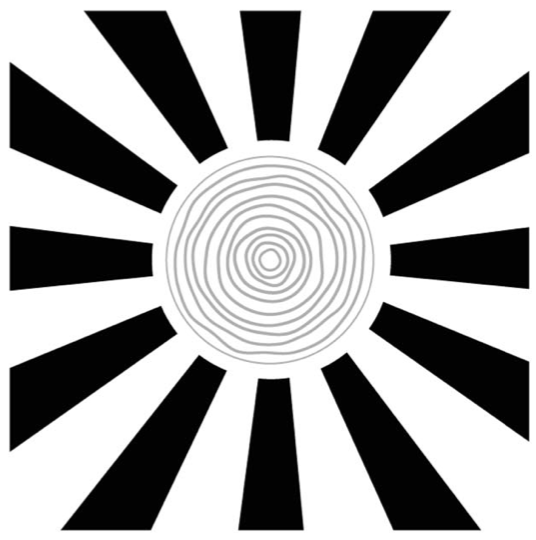}
		\caption{Billock and Tsou's phenomena: the presentation of funnel pattern in the centre induces an illusory perception of tunnel pattern in surround after a flickering (image on the \emph{left}). We have a reverse effect on the \emph{right}. Taken from \cite{billock2007}.}
		\label{fig:Billock-Tsou}
	\end{figure}

	Even more striking visual effects 
	have been obtained in the psychophysical experiments reported by \cite{billock2007}, see Fig.~\ref{fig:Billock-Tsou}. As in the case of the MacKay effect, they found that biasing stimuli could induce orthogonal responses in the visual cortex. 
	More precisely, a funnel pattern localised at the fovea (resp. in the periphery) with a background flicker induces the perception of a tunnel pattern in the periphery (resp. at the fovea).  
	The spatial interaction is localized, meaning the hallucination does not extend through the physical stimulus nor into empty non-flickering regions. 
	
	
	This work is concerned with a theoretical description of MacKay and Billock and Tsou's illusory phenomena in $\v1$.
	%
	This is achieved by studying the properties of the Amari-type equation \cite[Eq. (3)]{amari1977} describing the dynamics of the activity $a:\R_{+}\times\mathbb R^2\to \R$ on $\v1$:
	\begin{equation}\label{eq:NF-intro}\tag{NF}
		\frac{\partial a}{\partial t} = - a +\mu \omega\ast f(a) + I.
	\end{equation}
	Here, $\ast$ denotes the spatial convolution operation, $\omega:\mathbb R^2 \to \mathbb R$ is an interaction kernel modelling cortical connections in $\v1$, $f$ is a sigmoid non-linearity, and $\operatorname{I}:\mathbb R^2 \to \mathbb R$ is the cortical representation of the presented static visual stimulus, that is assumed to be time-independent. 
	Finally, $\mu>0$ is a parameter measuring the strength of intra-neuron connectivity. In this work, following \cite{tamekue2022}, we assume that the parameter $\mu$ is smaller than the threshold parameter $\mu_c$ where cortical patterns (e.g., funnels, tunnels, spirals, checkerboards, cobwebs, etc...) spontaneously emerge in $\v1$ (see, e.g.~\cite{ermentrout1979,bressloff2001}). 
	Neurophysiologically seeing, this corresponds to considering an unaltered state where no spontaneous hallucinations emerge.
	
	By an asymptotic analysis of the properties of \eqref{eq:NF-intro} we 
	describe why the after-image in the MacKay effect consists of illusory contours in the background of the physical visual stimulus. 
	In particular, the result we provide here implies that a motion in the after-image moves at a right angle to the stimulus pattern. This is because the static physical stimulus and the after-image in $\v1$ are superimposed horizontal and vertical stripes combined with the fact that the inverse retino-cortical map conserves this opponency in the retina. 
	
	Our main finding is that while the MacKay effect is essentially a linear phenomenon, Billock and Tsou's experiments are completely non-linear phenomena that strongly depend on the shape of the non-linear function used to model the neuronal response after an activation.
	Moreover, due to the equivariance of equation \eqref{eq:NF-intro} with respect to the plane Euclidean group $\mathbf{E}(2)$, 
	we find that the MacKay effect results from the highly redundant information in the visual stimulus aiming to break its plane Euclidean symmetry. 
	The same is true for Billock and Tsou's phenomenon, where symmetry-breaking arises due to the localization of the visual stimulus in the visual field.

	We conclude this section by mentioning that, up to our knowledge, the only other attempt to describe these phenomena theoretically is due to \cite{nicks2021}. There, the authors study a different model of neuronal fields equation containing an adaptation variable and a feedback (state-dependent) external input. 
	Their theoretical result relies on bifurcation and multi-scale analysis, which can be applied only for values of $\mu$ near the threshold parameter $\mu_c$ and in the presence of fully distributed external inputs. In particular, their analysis does not apply to the range of $\mu$ that we consider, nor to localized inputs, such as those used by MacKay and Billock and Tsou. Nevertheless, they provide numerical results showing the capability of their model to reproduce Billock and Tsou's experiments.
	
	\subsubsection{Notation.}
	In the following, $d\in\{1,2\}$ is the dimension of $\R^d$ and $|x|$ denote the Euclidean norm of $x\in\R^d$. For $p\in\{1,\infty\}$, $L^p(\R^d)$ is the Lebesgue space of class of real-valued measurable functions $u$ on $\R^d$ such that $|u|$ is integrable over $\R^d$ if $p=1$, and $|u|$ is essentially bounded over $\R^d$ when $p=\infty$. We endow these spaces with their standard norms $\|u\|_1 = \int_{\R^d}|u(x)|dx$  and $\|u\|_\infty = \operatorname{ess}\sup_{x\in\R^d}|u(x)|$. We let $\cS(\R^d)$ be the Schwartz space of rapidly-decreasing $C^\infty(\R^d)$ functions,
	and $\cS'(\R^d)$ be its dual space, i.e., the space of tempered distributions. Then, $\cS(\R^d)\subset L^p(\R^d)$ and $L^p(\R^d)\subset\cS'(\R^d)$ continuously. 
	The Fourier transform of $u\in L^1(\R^2)$ is defined by
	\begin{equation}\label{eq::Fourier transform in S}
		\widehat{u}(\xi):= \cF\{u\}(\xi)=\int_{\R^d} u(x)e^{-2\pi i\langle x,\xi\rangle}dx,\quad\forall\xi\in\R^d.
	\end{equation}
	Since $\cS(\R^d)\subset L^1(\R^2)$, one can extend the above by duality to $\cS'(\R^d)$, and in particular to $L^\infty(\mathbb{R}^d)$.
	Finally the convolution of $u\in L^1(\R^d)$ and $v\in L^p(\R^d)$, $p\in\{1,\infty\}$, is 
	\begin{equation}\label{eq::spatial convolution}
		(u\ast v)(x) = \int_{\R^d}u(x-y)v(y)dy,\qquad\forall x\in\R^d.
	\end{equation}

	\section{Neuronal fields equations}
	%
	%
	%
	%
	%
	%
	%
	%
	In their seminal paper, \cite{ermentrout1979} develop a theory describing (spontaneous) geometric visual hallucinations perceived in the retina. More precisely, using bifurcation techniques near a static Turing-like instability, they found that a simplified biological model of neuronal fields equation suffices to describe the (spontaneous) formation of cortical patterns (horizontal, vertical and oblique stripes, square, hexagonal and rectangular patterns etc.) in $\v1$. Then, applying the inverse retino-cortical map to these patterns, they obtained some of the geometric visual hallucinations or form constants that \cite{kluver1966} had meticulously classified. We refer to Fig.~\ref{fig::funnel} for a visual illustration concerning funnel patterns. In their considerations, $\v1$ is treated as a sheet of isotropically interconnected excitatory and inhibitory neurons. A more biologically realistic model of neuronal fields, including the anisotropic properties of cortical connections in $\v1$ (orientation preference of ``simple'' cells, see \cite{hubel1959}), was done in \cite{bressloff2001}. The authors were then able to describe all of Klüver's form constants. 
	
	Due to the success of the Ermentrout and Cowan model in describing simple patterns, such as funnel patterns, we expect that a similar model (i.e. without orientations preference) should be sufficient to describe sensory hallucinations (visual illusions) induced by these patterns. We, therefore, consider in this work that neuronal activity in $\v1$ evolves according to equation \eqref{eq:NF-intro}.  It models the average membrane potential $a(x,t)$ of a neuron located at $x\in\R^2$ at time $t\ge0$. In the next section, we present the assumptions on the parameters involved in this equation.
	\subsection{Assumptions on model parameters}\label{ss::Assumptions on model parameters}
	Throughout the following we assume the response function $f$ to be an odd non-decreasing function of class $C^2(\R)$ such that $f'(0) = \max_{s\in\R} f'(s) = 1$. 
	Unless explicitly stated otherwise, $f$ is a nonlinear sigmoidal function.
	
	The kernel $\omega$ is taken to be a DoG distribution (difference of Gaussians, also called ``Mexican hat'' distribution). Namely, we let for all $x\in\R^2$
	\begin{equation}\label{eq::connectivity}
		\omega(x) = [2\pi\sigma^2]^{-1}e^{-\frac{|x|^2}{2\sigma^2}}-[2\pi\kappa^2\sigma^2]^{-1}e^{-\frac{|x|^2}{2\kappa^2\sigma^2}},
	\end{equation}	
	where $\kappa> 1$ and $0<\sigma<1$.
	Clearly, $\omega$ is radial and $\omega$ belongs to the Schwartz space $\cS(\R^2)$. Moreover, its Fourier transform is explicitly given by
	\begin{equation}\label{eq::Fourier transform of the kernel omega}
		\widehat{\omega}(\xi)= e^{-2\pi^2\sigma^2|\xi|^2}-e^{-2\pi^2\sigma^2\kappa^2|\xi|^2},\qquad\forall\xi\in\R^2,
	\end{equation}	
	and $\widehat{\omega}$ reaches its maximum at every vector $\xi_c\in\R^2$ such that $|\xi_c| = \sqrt{\log\kappa/\pi^2\sigma^2(\kappa^2-1)}=:q_c$. 
	
	Observe that with this choice, we fall into the framework of \cite{bressloff2001}, i.e., there exists a critical interaction parameter $\mu_c:=\widehat{\omega}(\xi_c)^{-1}$ around which spontaneous cortical patterns in $\v1$ emerge. 
	\begin{Rema}
		The kernel $\omega$ satisfies the \textit{balance}\footnote{For a homogeneous NF equation (i.e., if $I = 0$), this condition enforces the existence of a unique stationary state $a_0=0$ even if $f(0)\neq 0$. It was assumed, for instance, in \cite{nicks2021} for the derivation of the amplitude equation.} condition $\widehat{\omega}(0)=0$ between excitation and inhibition. Moreover, $\widehat{\omega}(\xi)\ge 0$ for all $\xi\in\R^2$ and therefore $\|\widehat{\omega}\|_\infty = \widehat{\omega}(\xi_c)$. Nevertheless, this condition is just for mathematical convenience since it is not explicitly required in our study. Indeed, the following kernel $\omega$ works as well
		\begin{equation}\label{eq::connectivity-autre}
			\omega(x) = [2\pi\sigma_1^2]^{-1}e^{-\frac{|x|^2}{2\sigma_1^2}}-\kappa[2\pi\sigma_2^2]^{-1}e^{-\frac{|x|^2}{2\sigma_2^2}},\quad x\in\R^2,
		\end{equation}
		where $\kappa\ge 1$, $0<\sigma_1<\sigma_2$ and $\sigma_1\sqrt{\kappa}<\sigma_2$.
	\end{Rema}

	\subsection{Mathematical preliminaries}\label{ss::Mathematical preliminaries}
	
	We briefly recall some useful results related to equation \eqref{eq:NF-intro}. It is straightforward to show that the r.h.s.~of equation \eqref{eq:NF-intro} is a Lipschitz continuous map on $L^\infty_t(\mathbb R)\times L^\infty_x(\mathbb R^2)$. Thus, it is standard to obtain that, for every external input $I\in L^\infty(\R^2)$ and any initial datum $a_0\in L^\infty(\mathbb R^2)$, equation \eqref{eq:NF-intro} admits a unique solution $a\in C([0,+\infty); L^\infty(\mathbb R^2))$. Let us recall the following.
	\begin{Def}[Stationary state]\label{def::stationary state to WC equation}
		Let $a_0,\;I\in L^\infty(\R^2)$. A stationary state $a_I\in L^\infty(\R^2)$ to equation \eqref{eq:NF-intro} is  a time-invariant solution, viz.
		\begin{equation}\label{eq::SS}
			a_I = \mu\omega\ast f(a_I)+I.
		\end{equation}
	\end{Def}
	
	
	Via the contraction mapping principle, one obtains  the existence of a unique stationary solution whenever  $\mu<\mu_0$, see \cite[Proposition 1]{tamekue2022}. Here, we let 
	\begin{equation}
		\mu_0 := \|\omega\|_1^{-1}\le\mu_c.
	\end{equation}
	This implies in particular that if $\mu<\mu_0$, the map $\Psi: L^\infty(\mathbb R^2)\to L^\infty(\mathbb R^2)$ associating to each external input $I$ its corresponding stationary state is well-defined and bi-Lipschitz continuous. 
	Observe that $\Psi$ is defined by
	\begin{equation}\label{eq::map Psi}
		\Psi(I) = I+\mu\omega\ast f(\Psi(I)),\quad\forall I\in L^\infty(\mathbb R^2).
	\end{equation}
	It is then immediate that $\Psi$ and $\Psi^{-1}$ are $\mathbf{E}(2)$-equivariant, 
	see e.g.~\cite[Appendix A]{ermentrout1979}. 
	\begin{Rema}\label{rmk::equivariance of Psi}
		As a consequence of the $\mathbf{E}(2)$-equivariance of $\Psi$, a subgroup $\Gamma\subset\mathbf{E}(2)$ is a symmetry group of the external input $I\in L^\infty(\R^2)$ if and only if it is a symmetry group of the output stationary state $\Psi(I)$. E.g., if $I(x) = I(x_1)$ then $\Psi(I)(x) = \Psi(I)(x_1)$, for all $x = (x_1,x_2)\in\R^2$.
	\end{Rema}
	
	\subsection{Binary representation of patterns}\label{s::Strategy: Binary pattern}
	
	Due to the retino-cortical map, funnel, and tunnel patterns are respectively given in Cartesian coordinates $x:=(x_1,x_2)\in\R^2$ of $\v1$ by
	\begin{equation}\label{eq::funnel and tunnel patterns}
		\hspace{-0.3cm}  P_F(x) = \cos(2\pi\lambda x_2),\quad P_T(x) = \cos(2\pi\lambda x_1),\quad\lambda>0.
	\end{equation}
	This choice is motivated by analogy with the (spontaneous) geometric hallucinatory patterns described in \cite{ermentrout1979} and \cite{bressloff2001}.
	\begin{figure}
		\centering
		\includegraphics[width = .75\linewidth]{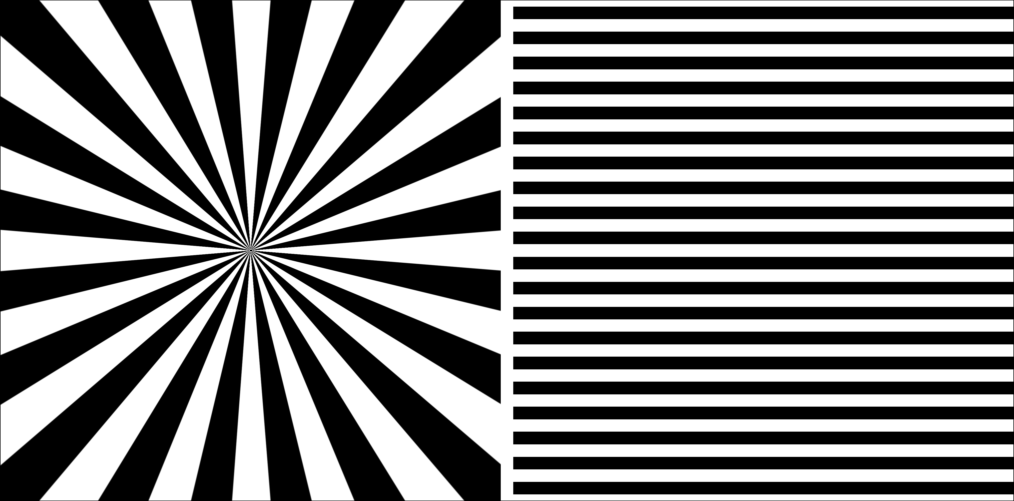}
		\caption{Funnel pattern with $\lambda=1$ in \eqref{eq::funnel and tunnel patterns}: Pattern in the retina (\emph{left}), corresponding one in $\v1$ (\emph{right}) after applying the retino-cortical map.}
		\label{fig::funnel}
	\end{figure}
	
	Given the above representation of funnel and tunnel patterns in cortical coordinates, to see how they look in terms of images, we represent them as contrasting white and black regions, see Fig~\ref{fig::funnel}. More precisely, define the binary pattern  $B_h$ of a function $h:\R^2\to\R$ by
	\begin{equation}
		B_h(x) = 
		\begin{cases}
			0, \quad \text{if } h(x)> 0 \quad\text{(black)}\\
			1, \quad \text{if } h(x)\le 0 \quad\text{(white)}.\\
		\end{cases}
	\end{equation}
	It follows that $B_h$ is essentially determined by the zero level-set of $h$.
	Since stimuli involved in the MacKay effect  and Billock and Tsou experiments are binary patterns, our strategy in describing these phenomena consists in characterising the zero level-set of output patterns. That is,
	we are mainly devoted to studying the qualitative properties of patterns by viewing them as binary patterns. 
	\section{MacKay effect}\label{s::MacKay effect}
	%
	%
	%
	%
	%
	%
	%
	In \cite[Theorem 1]{tamekue2022}, we proved that highly redundant information is  needed in the funnel and tunnel patterns for equation \eqref{eq:NF-intro} to reproduce the MacKay effect if $\mu<\mu_0/2$. More precisely, if the external input $I = P_F$ or $I = P_T$ in equation \eqref{eq:NF-intro}, $I$ and $\Psi(I)$ have the same binary pattern, then the same geometric shape in terms of images. Moreover, in the case where the response function $f$ is linear\footnote{Observe that if $f$ is linear, the fact that $f'(0)=1$ implies that $f(s)=s$. This is non-restrictive since the slope of $f$ can always be factored in the parameter $\mu$.}, we obtain that $\Psi(I)$ is even proportional to $I = P_F$ and $I = P_T$. These correspond respectively to $\xi_0 = (0,\lambda)$ and $\xi_0 = (\lambda,0)$ in the following.
	\begin{prop}\label{thm::stationary input in linear regime}
		Assume that $I\in L^\infty(\R^2)$ is given by $I(\cdot) = \cos(2\pi\langle \xi_0,\cdot\rangle)$, for some $\xi_0\in\R^2$
		and that the response function $f$ is linear. Then, if $\mu<\mu_0$, it holds
		\begin{equation*}
			a(\cdot,t)\xrightarrow[t\to\infty]{} \frac{I(\cdot)}{1-\mu\widehat{\omega}(\xi_0)},\quad\mbox{exponentially in}\quad L^\infty(\R^2).
		\end{equation*}
	\end{prop}
	\begin{pf} 
		The statement is a direct consequence of the fact that
		$I(\cdot) = \cos(2\pi\langle \xi_0,\cdot\rangle)$ satisfies $\omega\ast I=\widehat{\omega}(\xi_0)I$, and of the exponential convergence of $a$ to the stationary state. 
	\end{pf}
	We deduce that the system's resulting symmetry (the underlying Euclidean symmetry of the interaction kernel) restricts the geometrical shape of visual stimuli that can induce an illusory perception in the after-image. So in describing the MacKay effect with the considered model of neuronal activity \eqref{eq:NF-intro} (provided we are far from the bifurcation point, i.e. $\mu<\mu_0\le\mu_c$), it is necessary to break the Euclidean symmetry of the funnel and tunnel patterns by localising redundant information. 
	
	Concerning, e.g. the description of the MacKay effect related to funnel pattern, the ``MacKay rays'' (see the image on the left of Fig~\ref{fig:mackay} ), we use the following cortical representation:
	\begin{equation}\label{eq::analytical MacKay rays}
		I(x) = P_F(x)+\varepsilon H(-x_1) = \cos(2\pi\lambda x_2)+\varepsilon H(-x_1).
	\end{equation}
	Here $\varepsilon>0$, and $H$ is the Heaviside step function which would model redundant information in the funnel pattern. 
	
	The following result shows that equation \eqref{eq:NF-intro} with a linear response function $f$ suffices to describe illusory contours perceived in the after-image induced by the ``MacKay rays'' having \eqref{eq::analytical MacKay rays} as a $\v1$ analytical representation. 
	
	For simplicity, we assume in the rest of this section that parameters in the kernel $\omega$ defined in \eqref{eq::connectivity} are such that $2\pi^2\sigma^2 = 1$ and $\kappa^2 = 2$. We also choose $\mu := 1<2 = \mu_0$. 
	\begin{thm}\label{thm::stationary state for MacKay rays}
		Assume the response function $f$ is linear and the input $I$ is given by \eqref{eq::analytical MacKay rays}. Then, the unique stationary state to equation \eqref{eq:NF-intro} is given for all $(x_1,x_2)\in\R^2$ by
		\begin{equation}\label{eq::stationary state for MacKay rays}
			a_I(x_1,x_2) = \frac{\cos(2\pi\lambda x_2)}{1-\widehat{\omega}(\xi_0)}+\varepsilon g(x_1),\quad\xi_0:=(0,\lambda).
		\end{equation}
		Here $g:\R\to\R$ has a discrete set of zeroes on $(0,+\infty)$.
	\end{thm}
	The hypothesis on $f$ implies that the stationary equation \eqref{eq::SS} is linear. It follows that the first term in the r.h.s. of \eqref{eq::stationary state for MacKay rays} is the stationary state associated with the input $P_F$ provided by Proposition~\ref{thm::stationary input in linear regime} and $g$ is the stationary state associated to the second term in the r.h.s. of \eqref{eq::analytical MacKay rays}. Therefore, $g$ is the solution of the $1$-D stationary equation
	\begin{equation}\label{eq::1D ss}
		b(x) = H(-x)+(\omega_1\ast b)(x),\qquad x\in\R.
	\end{equation}
	where the $1$-D kernel $\omega_1$ is given for all $x\in\R$ by
	\begin{equation}\label{eq::connectivity 1-D}
		\omega_1(x) = [\sigma\sqrt{2\pi}]^{-1}e^{-\frac{x^2}{2\sigma^2}}-[\kappa\sigma\sqrt{2\pi}]^{-1}e^{-\frac{x^2}{2\kappa^2\sigma^2}}.
	\end{equation}
	Consequently, Theorem~\ref{thm::stationary state for MacKay rays} follows from the following.
	\begin{prop}\label{pro::ss for Heaviside}
		The solution $b\in L^\infty(\R)$ of \eqref{eq::1D ss} is given, for $x>0$, by
		\begin{gather}\label{eq::sol heaviside}
			e^{\pi x\sqrt{\frac{2\pi}{3}}}b(x) =\frac{\sqrt{3}}{\pi}\cos\left(\frac{\pi}{3}+\pi x\sqrt{\frac{2\pi}{3}}\right)+\frac{R(x)}{\pi x},\\
			\nonumber
			\text{where} \quad  |R(x)|\le\frac{2}{\pi\sqrt{10\pi}}. 
		\end{gather}
		Moreover, letting $(\theta_k)_{k}$ and $(\tau_k)_{k}$ be respectively zeroes and extrema of $x\mapsto \cos(\pi/3+\pi x\sqrt{2\pi/3})$ for $x>0$, the zeroes of $b$ in $(0,+\infty)$ are a countable sequence $(\rho_k)_{k}$  such that $\rho_k$ is unique in the interval $J_k:=]\tau_k,\tau_{k+1}[$ for all $k\in\N^{*}$ and 
		\begin{equation}\label{eq::asymptotic of zeroes of b}
			|\theta_{k+1}-\rho_k|\le\frac{\sqrt{3}}{\pi\sqrt{2\pi}}\arcsin\left(\frac{2}{\pi(3k-1)\sqrt{5}}\right).
		\end{equation}
	\end{prop}
	The proof of Proposition~\ref{pro::ss for Heaviside} relies on harmonic and complex analysis techniques. Due to space constraints, we only present a sketch of the proof in the following. We refer to a work in preparation for the complete details.
	
		\textbf{Proof of Proposition~\ref{pro::ss for Heaviside}.} Set $I(x) := H(-x)$ in \eqref{eq::1D ss}. Applying the Fourier transform in $\cS'(\R)$ to both sides of \eqref{eq::1D ss}, we obtain
		\begin{equation}\label{eq::SS Fourier DoG K}
			\widehat{b}(\xi) = (1+\widehat{K}(\xi))\widehat{I}(\xi),\qquad\qquad\xi\in\R,
		\end{equation}
		where we have set
		\begin{equation}
			\widehat{K}(\xi) := \frac{\widehat{\omega_1}(\xi)}{1-\widehat{\omega_1}(\xi)}.
		\end{equation}
		Obviously one has $\widehat{K}\in\cS(\R)$ so that its inverse Fourier transform $K\in\cS(\R)$ can be computed for all $x\in\R$ by
		\begin{equation}\label{eq:: K}
			K(x) = \int_{-\infty}^{+\infty}e^{2i\pi\xi x}\widehat{K}(\xi) d\xi = \int_{-\infty}^{+\infty}\frac{e^{2i\pi\xi x}\widehat{\omega_1}(\xi)}{1-\widehat{\omega_1}(\xi)}d\xi.
		\end{equation}
		By a precise analysis on $\widehat{K}$, applying the residue Theorem, we can show the existence of a function $S\in L^\infty(\R)$ such that for all $x\in\R\setminus\{0\}$, it holds
		\begin{equation}\label{eq:dev0}
			\frac{e^{\pi |x| \sqrt{\frac{2\pi}{3}}}K(x)}{2\sqrt{\pi}}=\cos\left(\frac{\pi}{12}+\pi |x|\sqrt{\frac{2\pi}{3}}\right)+\frac{S(x)}{|x|},
		\end{equation}
		\begin{equation}\label{eq:reste0}
			\vert S(x)\vert\leq \frac 2{\pi\sqrt{6\pi}},\qquad\forall x\in\R\setminus\{0\}.
		\end{equation}
		Taking now the inverse Fourier transform of equation \eqref{eq::SS Fourier DoG K} in the space $\cS'(\R)$, we find
		\begin{equation}\label{eq::explicit b concerning I}
			b(x) = I(x)+(K\ast I)(x),\qquad x\in\R\setminus\{0\}.
		\end{equation}
		Letting $I(x) = H(-x)$, we obtain \eqref{eq::sol heaviside}. 
		Standard arguments based on the intermediate value Theorem, allow us to prove the second part of the statement.
	\begin{Rema}
		Theorem~\ref{thm::stationary state for MacKay rays} implies that if the external input is the $\v1$ representation of the ``MacKay rays'' defined by \eqref{eq::analytical MacKay rays}, then the associated stationary state corresponds to the $\v1$ representation of the after-image reported by \cite{mackay1957}. Moreover, we have the exponential convergence of $a(\cdot,t)$ on the stationary state when $t\to\infty$. It follows that equation \eqref{eq:NF-intro} theoretically describes the MacKay effect associated with the ``MacKay rays'' at the cortical level. Due to the retino-cortical map, we deduce the theoretical description of the MacKay effect for the ``MacKay rays'' in the retina.
		
		
		We refer to Section~\ref{s::Numerical experiments} for numerical results.
	\end{Rema}
	\begin{Rema}
		One can prove that no illusory contours are present in the after-image using a Gaussian kernel to model synaptic interactions (i.e., discarding the excitatory/inhibitory nature of interactions).
		Indeed, the corresponding kernel $K$ in \eqref{eq:dev0} is positive on $\R\setminus\{0\}$ and thus letting $I(\cdot) = H(-\cdot)$ in \eqref{eq::explicit b concerning I} shows that $b$ is non-negative.
	\end{Rema}
	\section{Billock and Tsou's experiments}\label{s::Billock and Tsou's experiments} 
	In \cite{tamekue2022}, we exhibited numerical results showing the capability of equation \eqref{eq:NF-intro} to reproduce Billock and Tsou's experiments. The stimuli used in these experiments are funnel or tunnel patterns localised at the fovea or periphery. Due to the retino-cortical map, this corresponds to taking as external inputs in equation \eqref{eq:NF-intro}, $I= \varepsilon P_Fv$ or $I=\varepsilon P_Tv$, where $\varepsilon>0$ and $v$ is a localised function either in the left or in the right area of the cortex. In particular, the external inputs in these experiments do not fill all of the visual field. The function $v$ can then be thought of as a localised control aiming to break the \textit{global} plane Euclidean symmetry of stimuli patterns. 
	
	In this section, we prove that the equation \eqref{eq:NF-intro} with a linear response function $f$ cannot describe these phenomena: In contrast to the MacKay effect, the phenomena reported by Billock and Tsou are completely nonlinear. We mention that the numerical experiments of Section~\ref{s::Numerical experiments} will also highlight that the shape of the nonlinearity is crucial and, in particular, they suggest that an asymmetric nonlinearity is essential for the description. 
	
	We will focus on the  funnel pattern localised at the fovea. In $\v1$, it corresponds to the following external input.
	\begin{equation}\label{eq::Billock and Tsou input}
		I(x) =  \cos(2\pi\lambda x_2)H(-x_1),\qquad\lambda>0,\quad x\in\R^2,
	\end{equation}
	where $H$ is the Heaviside step function. For ease of notation, we assume the kernel $\omega$ in \eqref{eq::connectivity} is such that $\kappa^2=2$. If $\delta$ is the Dirac distribution at zero and $h\in\cS'(\R^2)$, 
	$$
	h_{\ast}^0 := \delta,\quad h_{\ast}^j:=h\ast h\ast\cdots\ast h,\qquad j\in\N^{*}.
	$$
	By Newton binomial formula, for all $h, g\in\cS'(\R^2)$, $n\in\N^{*}$
	\begin{equation}\label{eq::Newton binomial formula for convolution fo functions}
		(h+g)_{\ast}^n = \sum\limits_{j=0}^n\left(\begin{array}{c}
			n\\j
		\end{array}\right)h_{\ast}^{n-j}\ast g_{\ast}^j.
	\end{equation}
	Since the convolution of Gaussians with zero mean remains a Gaussian with zero mean, the following is a direct consequence of \eqref{eq::Newton binomial formula for convolution fo functions}.
	\begin{lem}\label{lem::general solution}
		Let $I\in\cS'(\R^2)$. Assume that the response function $f$ is linear. If $\mu<\mu_0$, the stationary state $a_I\in\cS'(\R^2)$ of \eqref{eq:NF-intro} is given by
		\begin{equation}\label{eq::general stationary solution}
			a_I = I+\sum\limits_{n=1}^\infty\mu^n\sum\limits_{j=0}^n\left(\begin{array}{c}
				n\\j
			\end{array}\right)(-1)^jg_{n,j}\ast I.
		\end{equation} 
		Here $g_{n,j}$ is the Gaussian defined for $x\in\R^2$, $n\in\N^{*}$ by
		\begin{equation}\label{eq::gaussian n and j}
			g_{n,j}(x) = \frac{1}{2\pi(n+j)\sigma^2} e^{-\displaystyle\frac{|x|^2}{2(n+j)\sigma^2}},\quad j\in[|0,n|].
		\end{equation}
	\end{lem}
	\begin{prop}\label{pro::stationary state Billock and Tsou}
		Assume that the response function $f$ is linear and  $I\in L^\infty(\R^2)$ is defined by \eqref{eq::Billock and Tsou input}. If $\mu<\mu_0$, there exists a function $R\in L^\infty(\R)$ such that the stationary state of \eqref{eq:NF-intro} is given for all $x\in\R^2$, by
		\begin{equation}
			a_I(x) = [H(-x_1)+R(x_1)]\cos(2\pi\lambda x_2).
		\end{equation}
	\end{prop}
	\begin{pf}
		By Lemma~\ref{lem::general solution}, one computes for all $x\in\R^2$,
		\begin{eqnarray}
			(g_{n,j}\ast I)(x)&=& \frac{\cos(2\pi\lambda x_2)}{\sigma\sqrt{2\pi(n+j)}} e^{-\frac{\lambda^2(n+j)\sigma^2}{2}}\int_{-\infty}^{0} e^{-\frac{(x_1-y_1)^2}{2(n+j)\sigma^2}}dy_1\nonumber\\
			&=&\frac{e^{-\frac{\lambda^2(n+j)\sigma^2}{2}}}{2}\erfc\left(\frac{x_1}{\sigma\sqrt{2(n+j)}}\right)\cos(2\pi\lambda x_2),\nonumber
		\end{eqnarray}
		where $\erfc$ is the complementary error function. 
	\end{pf}
	\begin{Rema}
		Proposition~\ref{pro::stationary state Billock and Tsou} shows that the output $a_I$ associated with \eqref{eq::Billock and Tsou input} has a contribution in the right area of the cortex given by
		\begin{equation}\label{eq::output in the right cortex}
			a_{I,r}(x) = R(x_1)\cos(2\pi\lambda x_2),\qquad x_1>0.
		\end{equation}
		Since $a_{I,r}$ depends on the factor $\cos(2\pi\lambda x_2)$, the retino-cortical map tells us that a visual stimulus consisting of fan shapes in the centre induces an after-image containing fan shapes in the periphery instead of concentric rings only, as Billock and Tsou reported. Equation \eqref{eq:NF-intro} (provided we are far from the bifurcation point, i.e. $\mu<\mu_0\le\mu_c$) with a linear response function cannot describe Billock and Tsou's experiments. Therefore, these phenomena depend fundamentally on the presence of the nonlinearity $f$.
	\end{Rema}
	\section{Numerical results}\label{s::Numerical experiments} 
	
	The numerical implementation is performed with Julia and is available at \url{https://github.com/dprn/MacKay-Billock_Tsou-2022/}. Given an input $I$, the stationary state $a_I$ is numerically implemented via an iterative fixed-point method. The cortical data is defined on a square $(x_1,x_2)\in[-L, L]^2$, $L=10$ with steps $\Delta x_1 = \Delta x_2 = 0.01$. 
	
\begin{figure}
		\centering
		\includegraphics[width = .74\linewidth]{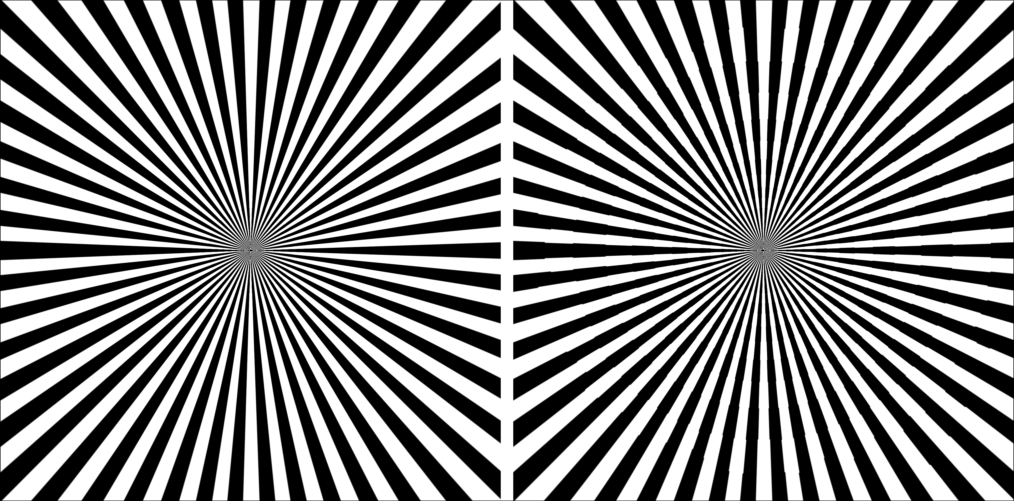}
		\caption{Initial stimulus (``MacKay rays'', \emph{left}) inducing the MacKay effect (\emph{right}).}
		\label{fig:mackay-funnel}
	\end{figure}

	For the reproduction of the MacKay effect, parameters in the kernel $\omega$ given by \eqref{eq::connectivity} are $2\pi^2\sigma^2 = 1$ and $\kappa^2=2$.  We exhibit in Fig.~\ref{fig:mackay-funnel} the MacKay effect for ``MacKay rays'' $I(x) = \cos(5\pi x_2)+\varepsilon H(2-x_1)$, $\varepsilon = 0.025$. Here, we use a linear response function ($f(s)=s$). We stress that the phenomenon can be reproduced with any odd sigmoidal function, see e.g.~\cite[Fig.~3]{tamekue2022}.
	
\begin{figure}
		\centering
		\includegraphics[width = .74\linewidth]{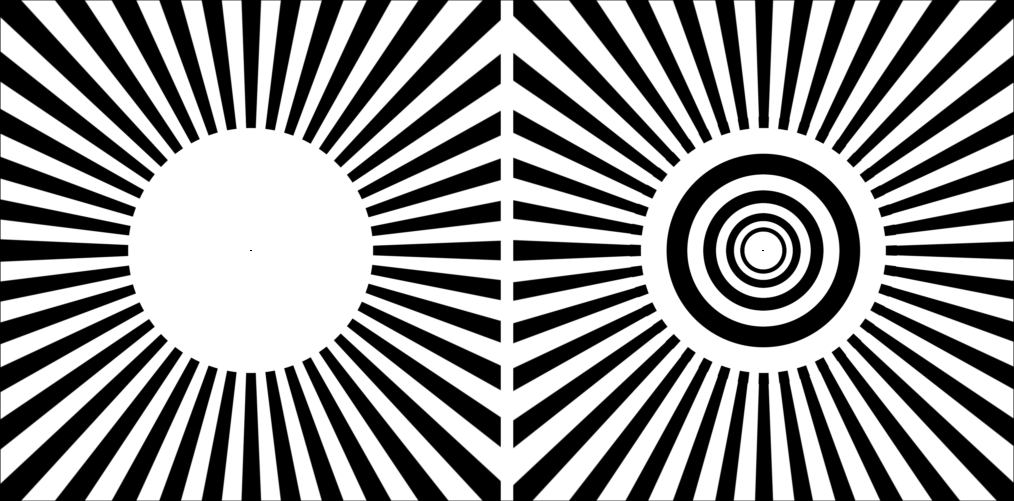}
		\caption{Billock and Tsou’s experiments: funnel stimulus localised at the periphery (\emph{left}) and after-image (\emph{right}).}
		\label{fig:Billock-funnel_surround}
	\end{figure}

\begin{figure}
		\centering
		\includegraphics[width = .74\linewidth]{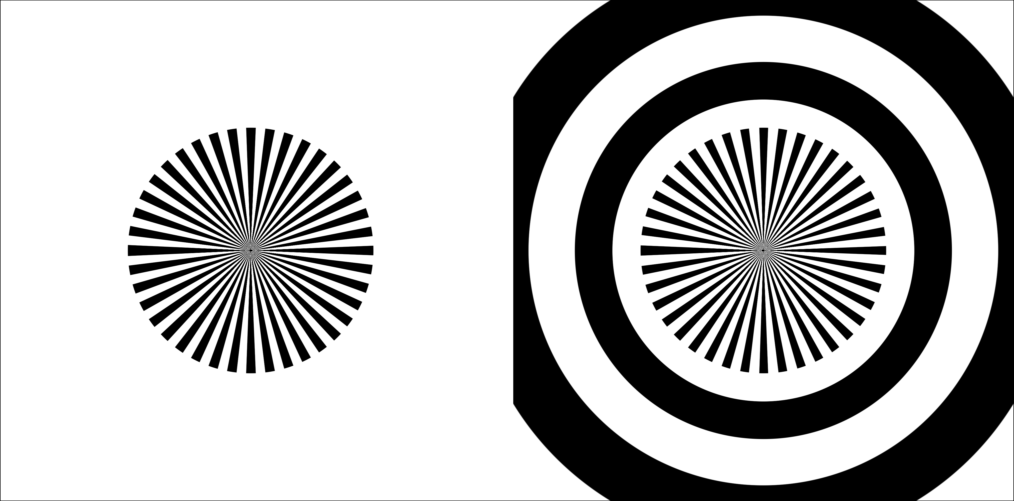}
		\caption{Billock and Tsou’s experiments: funnel stimulus localised at the fovea (\emph{left}) and after-image (\emph{right}).}
		\label{fig:Billock-funnel_fovea}
	\end{figure}

\begin{figure}
		\centering
		\includegraphics[width=.35\linewidth]{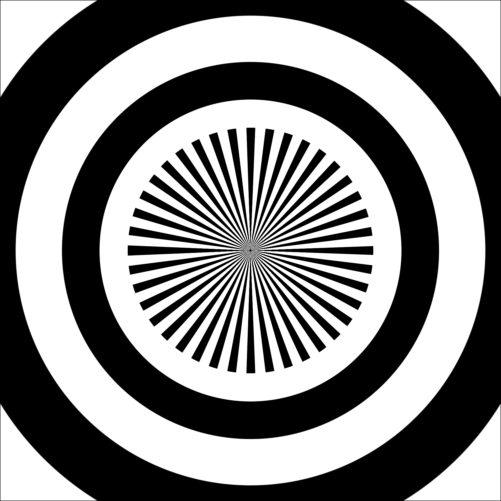}\hspace{1em}
		\includegraphics[width=.35\linewidth]{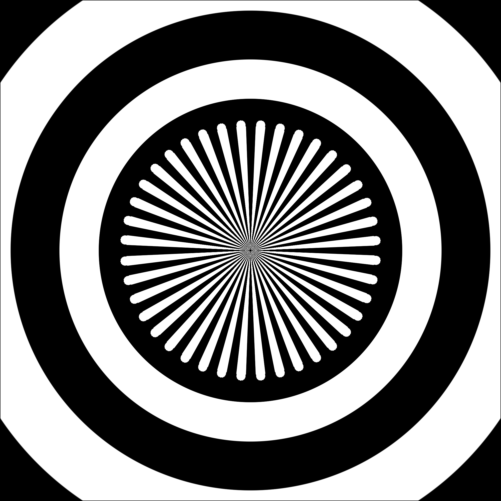}
		\caption{After-images in Billock and Tsou's experiments: funnel stimulus localised at the fovea as in Fig.~\ref{fig:Billock-funnel_fovea} (right). On the \emph{right} (resp.~\emph{left}) stimulus extends (resp.~does not) to the periphery.}
		\label{fig:billock-funnel}
	\end{figure}

	In \cite[Figs.~5 and 6]{tamekue2022}, we illustrated the capability of equation \eqref{eq:NF-intro} to reproduce Billock and Tsou experiments with the nonlinear response function $f(s) = (1+\exp(-s+0.25))^{-1}-(1+\exp(0.25))^{-1}$. In Section~\ref{s::Billock and Tsou's experiments}, we proved that a linear response function does not reproduce these phenomena. We exhibit in Figs.~\ref{fig:Billock-funnel_surround} and \ref{fig:Billock-funnel_fovea} Billock and Tsou's experiments for a funnel-like stimulus localised at the periphery and fovea, respectively. As images, we have a fan shape pattern at the periphery (resp. fovea) and white in the fovea (resp. periphery). We use the kernel $\omega$ defined in \eqref{eq::connectivity-autre} with $\sigma_1=0.1$, $\sigma_2=0.5$ and $\kappa=4.56$. In Fig.~\ref{fig:Billock-funnel_surround}, the stimulus is $I(x) = \cos(4\pi x_2)H(x_1-6)$, and the nonlinearity is $f(s) = \max(-0.2,\min(1,1.7s))$. In Fig.~\ref{fig:Billock-funnel_fovea}, the stimulus is $I(x) = \cos(4\pi x_2)H(6-x_1)$ and the nonlinearity is $f(s) = \max(-0.2,\min(1,1.2s))$. In Fig.~\ref{fig:billock-funnel}, the stimulus is $I(x) = \cos(4\pi x_2)H(6-x_1)$: On the left, the nonlinearity used is $f(s) = \max(-0.2,\min(1,1.2s))$, and on the right, we use $f(s) = \max(-1.2,\min(1,s))$. In the after-image on the left of Fig.~\ref{fig:billock-funnel}, the fan shape does not extend to the periphery, whereas on the right, the fan shape extends through the periphery. 
	
	To see how the shape of the nonlinearity is involved in the reproducibility, we consider the family of nonlinear functions $f_{m\alpha}(s) = \max(-m,\min(1,\alpha s))$. 
	We exhibit in Fig.~\ref{fig:Billock-heatmap} the range of parameters $(m,\alpha)\in \{ (k/10,\ell/10) \mid k = 0,\ldots,20, \, \ell = 1,\ldots, 20\}$ where equation \eqref{eq:NF-intro} with the nonlinear response function $f_{m\alpha}$ reproduces Billock and Tsou's experiments or not, for a funnel-like stimulus $I(x) = \cos(4\pi x_2)H(6-x_1)$. The \textit{magenta} (resp. \textit{black}) region corresponds to the value of $(m,\alpha)$ where $f_{m\alpha}$ reproduces (resp. do not reproduce) the phenomenon and the \textit{yellow} region corresponds to the value of $(m,\alpha)$ where $f_{m\alpha}$ reproduces the phenomenon, but the stimulus extends through the periphery. According to Billock and Tsou's experimental description, the good range of parameters $(m,\alpha)$ is that of the \emph{magenta} region.
	
	\section{Conclusions}
	We showed that breaking the Euclidean symmetry of stimuli is necessary for theoretically describing the MacKay effect and Billock and Tsou's experiments using equation \eqref{eq:NF-intro} whenever the parameter $\mu$ is smaller than $\mu_0$. To do this, we used a localised function. The issue of characterising all possible localized perturbations (i.e., ``controls'') of the funnel and tunnel inputs  that give rise to MacKay-type observations is still an open problem. 
	
	Although the Gaussian kernel is usually used in image processing and computer vision tasks due to its proximity to the visual system, it is unable to reproduce the phenomena described here. A physiological reason for this is that we used a one-layer model of the NF equation. It is not then biologically realistic to model synaptic interactions with a Gaussian, which would model only excitatory-type interactions between neurons. 
	
	Numerical results indicate that the anisotropic nature of cortical connections of ``simple'' cells in $\v1$ need not be integrated into equation \eqref{eq:NF-intro} to reproduce Billock and Tsou's experiments.  Moreover, the model reproduces the phenomena without a temporal flicker of the complementary region where the stimulus is not localized.
	
\begin{figure}
		\centering
		\includegraphics[width = .6\linewidth]{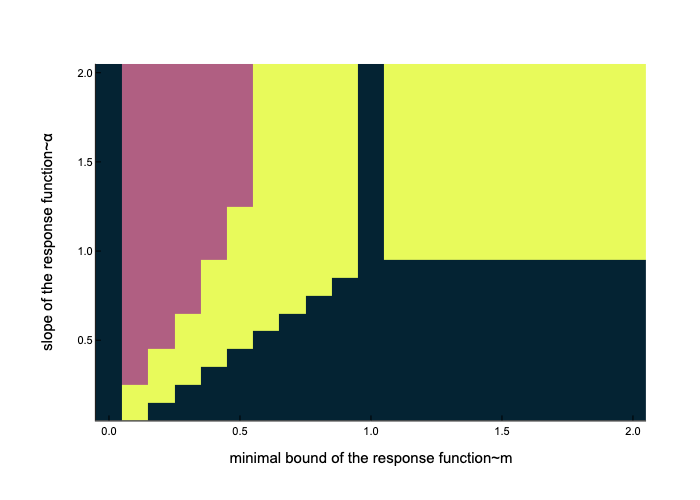}
		\caption{Range of $(m,\alpha)$ where $f_{m\alpha}$ reproduce Billock and Tsou's experiments or not, see text for details.}
		\label{fig:Billock-heatmap}
	\end{figure}

	Further work is ongoing to theoretically recover the numerical results in Fig.~\ref{fig:Billock-heatmap} for general nonlinear response functions. In particular, the numerical results suggest that to reproduce Billock and Tsou's phenomena, an asymmetric (i.e., not odd nor even) nonlinearity is required.
	
	\bibliography{tamekue_mackay-billock-tsou2022.bib}             

\begin{thebibliography}{11}
\providecommand{\natexlab}[1]{#1}
\providecommand{\url}[1]{\texttt{#1}}
\providecommand{\urlprefix}{URL }
\expandafter\ifx\csname urlstyle\endcsname\relax
  \providecommand{\doi}[1]{doi:\discretionary{}{}{}#1}\else
  \providecommand{\doi}{doi:\discretionary{}{}{}\begingroup
  \urlstyle{rm}\Url}\fi

\bibitem[{Amari(1977)}]{amari1977}
Amari, S. (1977).
\newblock Dynamics of pattern formation in lateral-inhibition type neural
  fields.
\newblock \emph{Biol. Cybern.}, 27(2), 77--87.

\bibitem[{Billock and Tsou(2007)}]{billock2007}
Billock, V.A. and Tsou, B.H. (2007).
\newblock Neural interactions between flicker-induced self-organized visual
  hallucinations and physical stimuli.
\newblock \emph{PNAS}, 104(20), 8490--8495.

\bibitem[{Bressloff et~al.(2001)Bressloff, Cowan, Golubitsky, Thomas, and
  Wiener}]{bressloff2001}
Bressloff, P.C., Cowan, J.D., Golubitsky, M., Thomas, P.J., and Wiener, M.C.
  (2001).
\newblock Geometric visual hallucinations, euclidean symmetry and the
  functional architecture of striate cortex.
\newblock \emph{Philos. Trans. R. Soc. Lond., B, Biol. Sci.}, 356(1407),
  299--330.

\bibitem[{Ermentrout and Cowan(1979)}]{ermentrout1979}
Ermentrout, G.B. and Cowan, J.D. (1979).
\newblock A mathematical theory of visual hallucination patterns.
\newblock \emph{Biol. Cybern.}, 34(3), 137--150.

\bibitem[{Helmholtz(1867)}]{helmholtz1867}
Helmholtz, H.L.F. (1867).
\newblock \emph{Optic physiologique}.
\newblock Masson.

\bibitem[{Hubel and Wiesel(1959)}]{hubel1959}
Hubel, D.H. and Wiesel, T.N. (1959).
\newblock Receptive fields of single neurones in the cat's striate cortex.
\newblock \emph{J. Physiol.}, 148(3), 574.

\bibitem[{Kl{\"u}ver(1966)}]{kluver1966}
Kl{\"u}ver, H. (1966).
\newblock Mescal and mechanisms of hallucinations.
\newblock \emph{Chicago: University of Chicago}.

\bibitem[{MacKay(1957)}]{mackay1957}
MacKay, D.M. (1957).
\newblock Moving visual images produced by regular stationary patterns.
\newblock \emph{Nature}, 180, 849--850.

\bibitem[{Nicks et~al.(2021)Nicks, Cocks, Avitabile, Johnston, and
  Coombes}]{nicks2021}
Nicks, R., Cocks, A., Avitabile, D., Johnston, A., and Coombes, S. (2021).
\newblock Understanding sensory induced hallucinations: From neural fields to
  amplitude equations.
\newblock \emph{SIAM J. Appl. Dyn. Syst.}, 20(4), 1683--1714.

\bibitem[{Tamekue et~al.(2022)Tamekue, Prandi, and Chitour}]{tamekue2022}
Tamekue, C., Prandi, D., and Chitour, Y. (2022).
\newblock Reproducing sensory induced hallucinations via neural fields.
\newblock In \emph{2022 IEEE-ICIP}, 3326--3330.

\bibitem[{Zeki et~al.(1993)Zeki, Watson, and Frackowiak}]{zeki1993}
Zeki, S., Watson, J.D., and Frackowiak, R.S. (1993).
\newblock Going beyond the information given: the relation of illusory visual
  motion to brain activity.
\newblock \emph{Proc. Royal Soc. B: Biol. Sci.}, 252(1335), 215--222.

\end{thebibliography}
	
	
	
	
	
	
	
	
\end{document}